\documentclass[twoside,twocolumn,9pt]{article}
\usepackage{extsizes}
\usepackage[super,sort&compress,comma]{natbib} 
\usepackage[version=3]{mhchem}
\usepackage[left=1.5cm, right=1.5cm, top=1.785cm, bottom=2.0cm]{geometry}
\usepackage{balance}
\usepackage{times,mathptmx}
\usepackage{sectsty}
\usepackage{graphicx} 
\usepackage{lastpage}
\usepackage[format=plain,justification=justified,singlelinecheck=false,font={stretch=1.125,small,sf},labelfont=bf,labelsep=space]{caption}
\usepackage{float}
\usepackage{fancyhdr}
\usepackage{fnpos}
\usepackage[english]{babel}
\addto{\captionsenglish}{%
  
}
\usepackage{array}
\usepackage{droidsans}
\usepackage{charter}
\usepackage[T1]{fontenc}
\usepackage[usenames,dvipsnames]{xcolor}
\usepackage{setspace}
\usepackage[compact]{titlesec}
\usepackage{hyperref}
\usepackage{epstopdf}%This line makes .eps figures into .pdf - please comment out if not required.

\definecolor{cream}{RGB}{222,217,201}

\begin{document}

\pagestyle{fancy}
\thispagestyle{plain}
\fancypagestyle{plain}{

%%%HEADER%%%
\fancyhead[C]{\includegraphics[width=18.5cm]{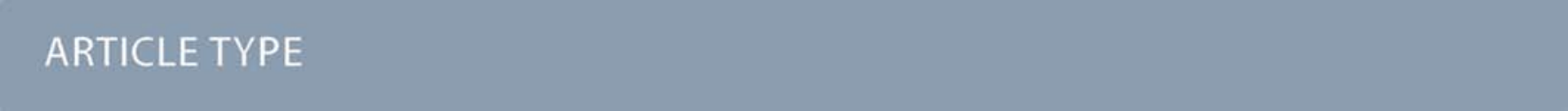}}
\fancyhead[L]{\hspace{0cm}\vspace{1.5cm}\includegraphics[height=30pt]{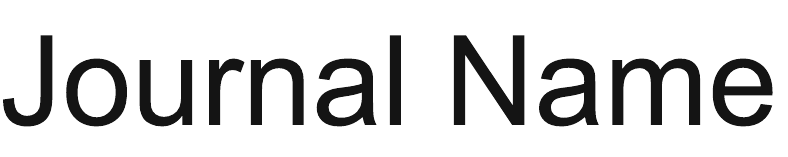}}
\fancyhead[R]{\hspace{0cm}\vspace{1.7cm}\includegraphics[height=55pt]{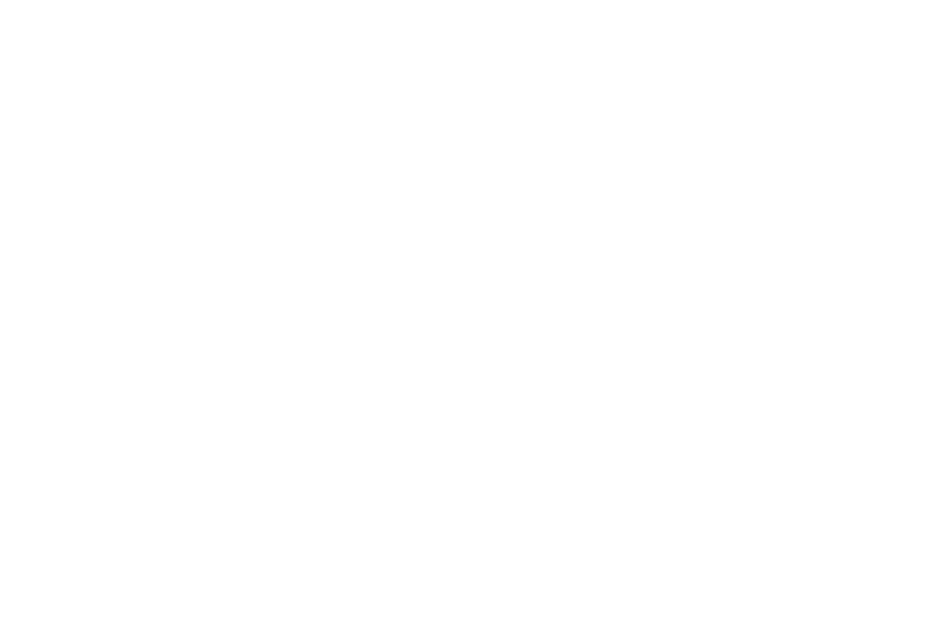}}
\renewcommand{\headrulewidth}{0pt}
}
%%%END OF HEADER%%%

%%%PAGE SETUP - Please do not change any commands within this section%%%
\makeFNbottom
\makeatletter
\renewcommand\LARGE{\@setfontsize\LARGE{15pt}{17}}
\renewcommand\Large{\@setfontsize\Large{12pt}{14}}
\renewcommand\large{\@setfontsize\large{10pt}{12}}
\renewcommand\footnotesize{\@setfontsize\footnotesize{7pt}{10}}
\makeatother

\renewcommand{\thefootnote}{\fnsymbol{footnote}}
\renewcommand\footnoterule{\vspace*{1pt}% 
\color{cream}\hrule width 3.5in height 0.4pt \color{black}\vspace*{5pt}} 
\setcounter{secnumdepth}{5}

\makeatletter 
\renewcommand\@biblabel[1]{#1}            
\renewcommand\@makefntext[1]% 
{\noindent\makebox[0pt][r]{\@thefnmark\,}#1}
\makeatother 
\renewcommand{\figurename}{\small{Fig.}~}
\sectionfont{\sffamily\Large}
\subsectionfont{\normalsize}
\subsubsectionfont{\bf}
\setstretch{1.125} %In particular, please do not alter this line.
\setlength{\skip\footins}{0.8cm}
\setlength{\footnotesep}{0.25cm}
\setlength{\jot}{10pt}
\titlespacing*{\section}{0pt}{4pt}{4pt}
\titlespacing*{\subsection}{0pt}{15pt}{1pt}
%%%END OF PAGE SETUP%%%

%%%FOOTER%%%
\fancyfoot{}
\fancyfoot[LO,RE]{\vspace{-7.1pt}\includegraphics[height=9pt]{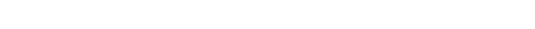}}
\fancyfoot[CO]{\vspace{-7.1pt}\hspace{13.2cm}\includegraphics{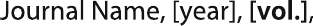}}
\fancyfoot[CE]{\vspace{-7.2pt}\hspace{-14.2cm}\includegraphics{headfoot/RF}}
\fancyfoot[RO]{\footnotesize{\sffamily{1--\pageref{LastPage} ~\textbar  \hspace{2pt}\thepage}}}
\fancyfoot[LE]{\footnotesize{\sffamily{\thepage~\textbar\hspace{3.45cm} 1--\pageref{LastPage}}}}
\fancyhead{}
\renewcommand{\headrulewidth}{0pt} 
\renewcommand{\footrulewidth}{0pt}
\setlength{\arrayrulewidth}{1pt}
\setlength{\columnsep}{6.5mm}
\setlength\bibsep{1pt}
%%%END OF FOOTER%%%

%%%FIGURE SETUP - please do not change any commands within this section%%%
\makeatletter 
\newlength{\figrulesep} 
\setlength{\figrulesep}{0.5\textfloatsep} 

\newcommand{\topfigrule}{\vspace*{-1pt}% 
\noindent{\color{cream}\rule[-\figrulesep]{\columnwidth}{1.5pt}} }

\newcommand{\botfigrule}{\vspace*{-2pt}% 
\noindent{\color{cream}\rule[\figrulesep]{\columnwidth}{1.5pt}} }

\newcommand{\dblfigrule}{\vspace*{-1pt}% 
\noindent{\color{cream}\rule[-\figrulesep]{\textwidth}{1.5pt}} }

\makeatother
%%%END OF FIGURE SETUP%%%

%%%TITLE, AUTHORS AND ABSTRACT%%%
\twocolumn[
  \begin{@twocolumnfalse}
\vspace{3cm}
\sffamily
\begin{tabular}{m{4.5cm} p{13.5cm} }

\includegraphics{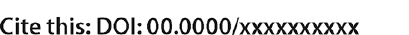} & \noindent\LARGE{\textbf{Structure of jammed ellipse packings with a wide range of aspect ratios}} \\%Article title goes here instead of the text "This is the title"
\vspace{0.3cm} & \vspace{0.3cm} \\

 & \noindent\large{Sebastian Rocks and Robert S. Hoy$^{\ast}$} \\%Author names go here instead of "Full name", etc.

\includegraphics{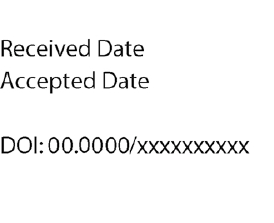} & \noindent\normalsize{ 
Motivated in part by the recent observation of liquid glass in suspensions of ellipsoidal colloids, we examine the structure of jammed ellipse packings over a much wider range of particle aspect ratios ($\alpha$) than has been previously attempted.  
We determine  $\phi_{\rm J}(\alpha)$  to high precision, and find empirical analytic formulae that predict $\phi_{\rm J}(\alpha)$ to within less than 0.1\% for all $1 \leq \alpha \leq 10$,
for three different particle dispersities.
Then we explore how these packings’ local structural order varies with $\alpha$.   
We find that the densest packings possess unusually-well-defined nearest-neighbor shells, including both a higher fraction $f_{\rm Z = 6}$ of particles with exactly six contacts and a previously-unreported short-range order marked by ``kinetically suppressed'' regions in their positional-orientational pair correlation function $g(r,\Delta \theta)$.
We also show that the previously-reported approach to isostaticity (coordination number $Z_{\rm J} \to Z_{\rm iso} \equiv 6$) with increasing $\alpha$ is interrupted and then reversed as local nematic order increases: $Z_{\rm J}(\alpha)$ drops towards 4 as ellipses are more often trapped by contacts with a parallel-oriented neighbor on either side and a perpendicularly-oriented neighbor on either end.  
Finally we show that  $\phi_{\rm J}/\phi_{\rm s}$ (where $\phi_{\rm s}$ is the saturated RSA packing density) is nearly $\alpha$-independent  for systems that do not develop substantial local hexatic or nematic order during compression.
 } \\

\end{tabular}

 \end{@twocolumnfalse} \vspace{0.6cm}

  ]
%%%END OF TITLE, AUTHORS AND ABSTRACT%%%

%%%FONT SETUP - please do not change any commands within this section
\renewcommand*\rmdefault{bch}\normalfont\upshape
\rmfamily
\section*{}
\vspace{-1cm}

%%%FOOTNOTES%%%

\footnotetext{Department of Physics, University of South Florida, Tampa, FL 33620, USA.}
\footnotetext{E-mail: rshoy@usf.edu}

%%%END OF FOOTNOTES%%%

\section{Introduction}
\label{sec:intro}

Most real granular materials are composed of aspherical, shape-anisotropic particles.
Theoretical efforts aiming to explain the various ways in which constituent-particle anisotropy affects systems' jamming phenomenology have focused primarily on simple models in which the degree of anisotropy can be controlled by varying one parameter: the aspect ratio $\alpha$.
The variation of jamming phenomenology with $\alpha$ is the simplest for high-symmetry convex shapes, and as a consequence, the theoretical study of anistropic-particle jamming began with ellipses and ellipsoids.\cite{donev04,donev07,delaney05}

Jamming of low-aspect-ratio ellipses has been extensively studied\cite{donev04,donev07,mailman09,delaney05,schreck10,vanderwerf18} and is now fairly well understood.
In particular, for $\alpha-1 \ll 1$, the linear increase in $\phi_{\rm J}$ [$\phi_{\rm J}(\alpha) - \phi_{\rm J}(1) \sim (\alpha-1)$] and the singularity in the average coordination number $Z_{\rm J}$ of marginally jammed states [$Z_{\rm J}(\alpha) - Z_{\rm J}(1) \propto \sqrt{\alpha - 1}$] have respectively been explained in terms of particles' ability to pack more efficiently than disks by rotating away from contacts\cite{donev04,donev07} and by the divergence in the number of quartic modes as $\alpha \to 1$.\cite{donev07,mailman09}
These features are closely associated with each other, in the sense that  $\phi_{\rm J}(\alpha) - \phi_{\rm J}(1) \sim [Z_{\rm J}(\alpha) - Z_{\rm J}(1)]^2$.
On the other hand, while these early studies explained the most essential features of the variation of  low-aspect-ratio ellipses' jamming phenomenology with $\alpha$, they did not establish precise analytic formulas for $\phi_{\rm J}(\alpha)$ or $Z_{\rm J}(\alpha)$, or examine the local structural ordering of jammed packings in much detail.

Recent experiments have demonstrated the existence of a ``liquid glass'' state in both quasi-2D\cite{zheng11,zheng14,mishra13} and 3D\cite{roller20, roller21} suspensions of ellipsoidal colloids.  
In this state, which occupies packing fractions $\phi$ that are between systems' orientational and translational glass transitions [i.e.\ all $\phi_{\rm g}^{\rm rot}(\alpha) \leq \phi \leq \phi_{\rm g}^{\rm trans}(\alpha)$], particles rotations' are arrested but they remain free to translate within  locally-nematic precursor domains.
The existence of this state was predicted nearly 25 years ago by mode coupling theory \cite{letz00} and confirmed nearly 10 years ago by  Monte Carlo simulations of hard ellipses \cite{zheng14}, but it remains poorly understood.
The well-established, intimate connection between the glass and jamming transitions\cite{liu98,charbonneau17} suggests that at least some of ellipses' liquid-glass state's physics is controlled by their jamming phenomenology.
However, jamming of ellipses with $\alpha$ that are sufficiently large for systems to form the (essential) locally-nematic precursor domains as systems are being compressed has been almost completely neglected by theorists.
Only Ref.\ \cite{delaney05} examined ellipses with $\alpha > 2.5$, and no studies have examined systems with $\alpha > 5$.

In this paper, we examine the structure of jammed ellipse packings over a much wider range of aspect ratios ($1 \leq \alpha \leq 10$) than has previously been attempted.\
All of our results for $\alpha <\sim 3$ are consistent with previous studies,\cite{donev04,donev07,mailman09,delaney05,schreck10,vanderwerf18} but we go beyond previous work by (1) identifying nearly-exact analytic expressions for $\phi_{\rm J}(\alpha)$ and (2) performing a detailed characterization of jammed states' local structural order.
We show that the primary signature distinguishing jammed ellipse packings with $\alpha \simeq \alpha_{\rm max}$ [where $\alpha_{\rm max}$ is the aspect ratio at which $\phi_{\rm J}(\alpha)$ is maximized] from those with lower $\phi_{\rm J}$ is that they possess unusually-well-defined nearest-neighbor shells, including both a higher fraction $f_{\rm Z = 6}$ of particles with exactly six contacts and a previously-unreported short-range order marked by ``kinetically suppressed''  regions in the positional-orientational pair correlation function $g(r,\Delta \theta)$.
For $\alpha > 3$, we show that $Z_{\rm J}$ drops slowly towards $4$ with increasing $\alpha$, as local nematic order increases and ellipses are more often trapped by contacts with a parallel-oriented neighbor on either side and a perpendicularly-oriented neighbor on either end.
This result stands in stark contrast to the one that might have been expected from Refs.\   \cite{donev04,donev07,mailman09,delaney05,schreck10,vanderwerf18}, which suggested $\lim_{\rm \alpha \to \infty} Z_{\rm J} = 6$.
We also show that the ratio $\phi_{\rm J}(\alpha)/\phi_{\rm s}(\alpha)$, where $\phi_{\rm s}(\alpha)$ is ellipses' random sequential adsorption (RSA) density, is nearly constant for systems that do not develop substantial local hexatic or nematic order during compression. 
Finally, by comparing results for three distinct particle dispersities, we show that all of the abovementioned results are general.

\section{Methods}
\label{sec:methods}

To facilitate comparison of jammed and saturated-RSA ellipse packings, we examined the same set of 81 different particle aspect ratios (over the range $1 \leq \alpha \leq 10$) considered in Ref.\ \cite{abritta22}.
Jammed ellipse packings were obtained using a Lubachevsky-Stillinger-like\cite{lubachevsky91} growth algorithm.
To understand the effects of particle dispersity, we employed three different probability distributions for the ellipses' inital minor-axis lengths $\sigma$:
\begin{equation} 
\begin{array}{c}
P_{\rm mono}(\sigma) = \delta(\sigma - .07 a) \\
\\
P_{\rm bi}(\sigma) = \displaystyle\frac{\delta(\sigma - .05 a)}{2} + \displaystyle\frac{\delta(\sigma - .07 a)}{2} \\
\\
P_{\rm contin}(\sigma) =  \Bigg{ \{ }  \begin{array}{ccc}
\displaystyle\frac{7}{4\sigma^2} & , & .05a \leq \sigma \leq .07a\\
\\
0 & , & \sigma < .05a\ \textrm{or}\ \sigma > .07a
\end{array}
\end{array} ,
\label{eq:dispersity} 
\end{equation}
where $\delta$ is the Dirac delta function and $a$ is an arbitrary unit of length.
$P_{\rm mono}$ yields monodisperse particles, $P_{\rm bi}$ yields the bidisperse 50:50 1:1.4 particles with radii $R_{\rm small} = .5{ a}$, $R_{\rm large} = .7{ a}$ that have been the standard model for studies of granular materials for the past 25 years,\cite{speedy98b,ohern03} and $P_{\rm contin}$ yields continuously-polydisperse systems in which equal areas are occupied by particles of different sizes.

For each $\alpha$ and particle dispersity $x$ [i.e.\ for each $P_{\rm x}(\sigma)$], 100 jammed packings were prepared using the following procedure:
$N = 1000$ nonoverlapping ellipses of aspect ratio $\alpha$ were placed with random positions and orientations in square $L \times L$ domains, with $L = 36.1818\sqrt{\alpha}a$.
Periodic boundary conditions were applied along both directions, so these initial states had packing fractions below $0.01$.
Jammed states were obtained using a Monte Carlo (MC) algorithm.
Each MC cycle consisted of: 
\begin{enumerate}
\item Attempting to translate particle $i$ by a random displacement of maximum magnitude $0.05fa$ along each Cartesian direction and rotate it  by an angle of maximum magnitude $(10f/\alpha)^\circ$,
\item Repeating step 1 for $i = 1,2,...,N$, and 
\item Increasing all particles' $\sigma$ by the maximum possible factor consistent with hard-particle constraints, i.e.\ the factor that brings one pair of ellipses into tangential contact.
\end{enumerate}
This implementation of step (3) preserved the particle dispersities defined in Eq.\ \ref{eq:dispersity}.
The move-size factor $f$ was set to $1$ at the beginning of the runs, and multiplied by $3/4$ whenever $100$ cycles had passed without a successful translation/rotation attempt.
Runs were terminated and the configurations were considered jammed when $f$ dropped below $10^{-9}$, the minimum value allowed by our double-precision numerical implementation of this algorithm.
Throughout this process, inter-ellipse overlaps were prevented using Zheng and Palffy-Muhoray's exact expression\cite{zheng07} for their distance of closest approach $d_{\rm cap}$.

We characterized the structural order of the jammed packings using  several commonly employed metrics:

In addition to $Z_{\rm J}$, we examined the fractions $f_{\rm Z = 6}$ ($f_{\rm Z = 4}$)  of particles that have exactly six (four) contacts.
$f_{\rm Z = 6} = 1$ in both the triangular lattice (the densest crystalline packing of both disks and ellipses, and  isostatic jammed ellipse packings, while $f_{\rm Z = 4} = 1$ in isostatic disk packings and in ``checkerboard''-like phases formed by perpendicularly-oriented, short single-layer lamellae.\cite{shah22}

Local nematic order was characterized using the standard order parameter
\begin{equation}
S = \displaystyle\frac{1}{18N} \displaystyle\sum_{i = 1}^N \displaystyle\sum_{j = 1}^{18} \displaystyle\frac{3 \cos^2(\Delta \theta_{ij}) \rangle -1}{2} \equiv \displaystyle\frac{3\langle \cos^2(\Delta \theta) \rangle -1}{2}  ,
\label{eq:nemS}
\end{equation}
where $\Delta \theta_{ij}$ is the orientation-angle difference between ellipses $i$ and $j$, and the average is performed over the 18 nearest neighbors of each ellipse.
Here 18 was chosen because it corresponds to the total number of first, second, and third nearest neighbors for particles in a triangular lattice; this choice makes $S$ a measure of \textit{mid-range} nematic order.
$S$ is $1$ for a perfectly-nematically-ordered and zero for an orientationally-disordered material.

Local hexatic order was characterized  using the Steinhardt-like\cite{steinhardt83} order parameter
\begin{equation}
\Psi_6 =  \displaystyle\frac{1}{6N} \displaystyle\sum_{i=1}^{N} \left| \displaystyle\sum_{j = 1}^{6} \exp(6i\Theta_{ij}) \right|.
\label{eq:psi6}
\end{equation}
Here $\Theta_{ij}$ is the angle between the vector $\vec{r}_{ij}$ connecting ellipses $i$ and $j$ and an arbitrary fixed axis, and the inner sum is taken over the 6 nearest neighbors of each monomer $i$.
This metric has been shown to be useful in identifying the onset of liquid-crystalline order in hard-disk systems.\cite{bernard11}
$\Psi_6$ is $1$ for the triangular lattice (at any density) since the angles between its $\{ \vec{r}_{ij} \}$ are multiples of $60^\circ$, and zero for a perfectly-orientationally-disordered material since the angles between its $\{ \vec{r}_{ij} \}$ are random.

To gain additional insight into the connections between variations in nematic and hexatic order and variations in $\phi_{\rm J}$, we examined the variance
\begin{equation}
\Sigma^2(R) = \langle n^2(R) \rangle - \langle n(R) \rangle^2
\label{eq:numvar}
\end{equation}
of the number of ellipses whose centers lie within randomly located circular ``windows'' of radius $R$.
The scaling of $\Sigma^2$ with $R$ is a sensitive measure of packings' ``uniformity''.\cite{torquato18b}
Crystals and quasicrystals have $\Sigma^2 \sim R^{d-1}$, standard amorphous packings have $\Sigma^2 \sim R^d$, and maximally random jammed (MRJ) packings have $\Sigma^2 \sim R^{d-1}\ln(R)$.\cite{torquato18b,torquato03}

Finally we calculated the positional-orientational pair correlation function $g(r,\Delta\theta)$, which is the ratio of the number of ellipse pairs with center-to-center distance $r$ and orientation-angle difference $\Delta\theta$ to the number that would be present in an ideal gas of these particles.
In other words $g(r,\Delta\theta)$ is just the generalization of the standard pair correlation function $g(r)$ to include orientation-angle differences. 
Our recent study\cite{abritta22} showed that this metric is key to understanding how the structure of saturated RSA ellipse packings varies with $\alpha$.

All numerical data presented below are averages over the 100 packings we prepared for each $\alpha$ and  $P_{\rm x}(\sigma)$.

\section{Results}
\label{sec:results}

\subsection{Basic features}

\begin{figure}[htbp]
\includegraphics[width=3.375in]{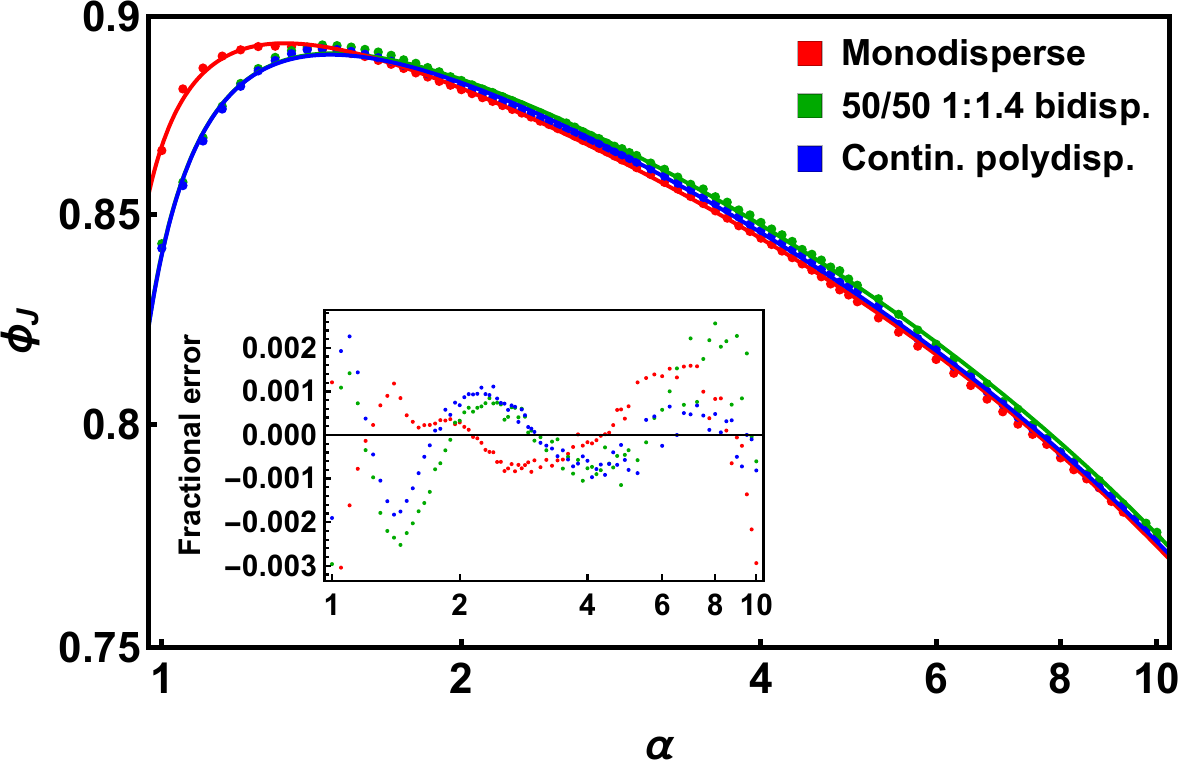}
\caption{Jamming densities for ellipses with $1 \leq \alpha \leq 10$.   Symbols show data from our LS runs while curves respectively show Eqs.\ \ref{eq:phiJmono}-\ref{eq:phiJcontin}, and the inset shows the fractional difference of the predictions of these equations from the data.}
\label{fig:phiJdata}
\end{figure}

Figure \ref{fig:phiJdata} shows $\phi_J(\alpha)$ for all three particle dispersities.
Differences between results for bidisperse and continuously-polydisperse systems are minimal, while the differences between these and results for monodisperse systems are expected from the latter's well-known tendency to crystallize even under rapid Lubachevsky-Stillinger-style compression.\cite{lubachevsky91}
All data for $\alpha <\sim 3$, and the basic features of the entire  $\phi_J(\alpha)$ curves, are qualitatively consistent with previous studies.\cite{donev04,donev07,delaney05,mailman09,schreck10,vanderwerf18}
Our data show that $\phi_{\rm J}(\alpha) > \phi_{\rm J,disks} \equiv  \phi_{\rm J}(1)$ for $1 < \alpha < 2.70$ ($1 < \alpha < 4.46$) [$1 < \alpha < 4.35$] for monodisperse (bidisperse) [continuously-polydisperse] ellipses, indicating that particle anisotropy enhances packability over these ranges of $\alpha$.
Surprisingly, bidisperse and continuously-polydisperse systems actually pack better (have a higher $\phi_J$) than monodisperse systems
for $\alpha >\sim 1.5$, suggesting that a size ratio of 1.4 is large enough for small ellipses to fill the gaps between larger ones in an at-least-semicoherent fashion.

We find that the $\phi_J$ for monodisperse, bidisperse, and continuously-polydisperse ellipses are respectively very well fit by
\begin{equation}
\phi_{\rm J}^{\rm mono}(\alpha) = \phi_{\rm J,disks}^{\rm mono}  \times \displaystyle\frac{1 + \frac{73}{120}\ln(\alpha) +  \frac{49}{9}(\alpha-1)}{1 + \frac{108}{19}(\alpha - 1) + \frac{13}{190}(\alpha-1)^2} ,
\label{eq:phiJmono}
\end{equation}
\begin{equation}
\phi_{\rm J}^{\rm bi}(\alpha) = \phi_{\rm J,disks}^{\rm bi}  \times \displaystyle\frac{1 + \frac{13}{20}\ln(\alpha) +  \frac{49}{10}(\alpha-1)}{1 + \frac{249}{50}(\alpha - 1) + \frac{5}{86}(\alpha-1)^2} ,
\label{eq:phiJbi}
\end{equation}
and
\begin{equation}
\phi_{\rm J}^{\rm contin}(\alpha) = \phi_{\rm J,disks}^{\rm contin}  \times \displaystyle\frac{1 + \frac{11}{16}\ln(\alpha) +  \frac{193}{40}(\alpha-1)}{1 + \frac{247}{50}(\alpha - 1) + \frac{10}{179}(\alpha-1)^2}.
\label{eq:phiJcontin}
\end{equation}
Here $\phi_{\rm J,disks}$ depends on both particle dispersity and the protocol with which jammed states are prepared.
For our bidisperse and continuously-polydisperse systems it takes on standard MRJ-like values, respectively $0.8404$ and $0.8402$.\cite{torquato00,ohern03}
For monodisperse systems it is substantially larger ($0.8669$)  owing to these systems' well-known tendency to crystallize even under rapid Lubachevsky-Stillinger-style compression \cite{lubachevsky91}.

The mean  fractional deviations of these expressions'  predictions from the ensemble-averaged measured $\phi_J$ are essentially zero, while the rms  fractional deviations, which are respectively $\sim 0.09\%$, $\sim 0.12\%$ and $0.09\%$ for monodisperse, bidisperse, and continuously-polydisperse ellipses, are only slightly above the lower bounds set by the statistical uncertainties on the measured $\phi_J$.
However, we do \textit{not} claim that any of Eqs.\ \ref{eq:phiJmono}-\ref{eq:phiJcontin} are exact expressions valid for all $\alpha$, or even that their functional form is the same as that of the ``true'' $\phi_{\rm J}^{\rm x}(\alpha)$ which could be obtained given infinite computer power.
We also emphasize that the coefficients preceding the $\ln(\alpha)$ and $(\alpha - 1)^x$ terms are preparation-protocol-dependent.

\begin{figure*}[htbp]
\includegraphics[width=7in]{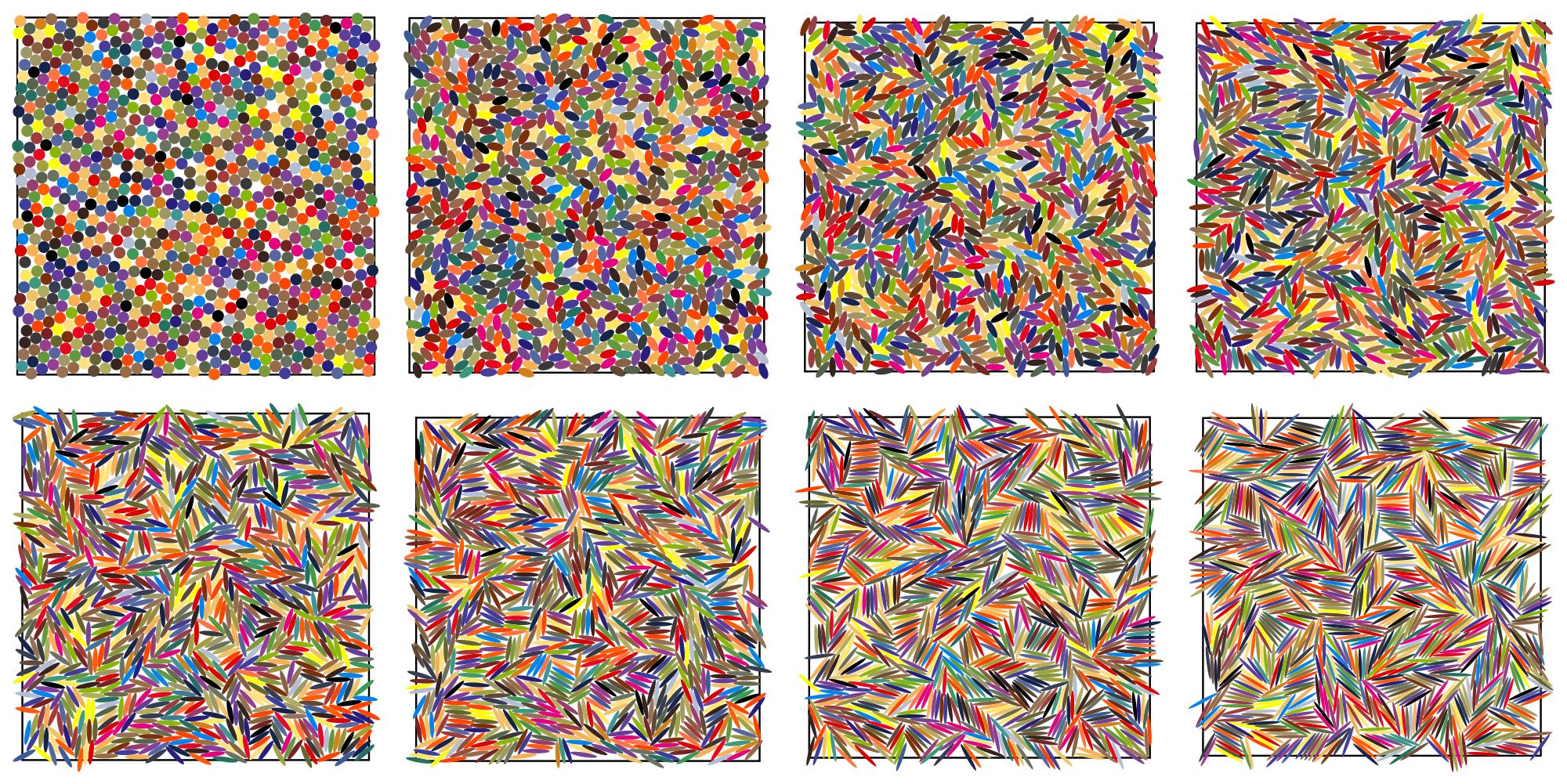}
\caption{Snapshots of jammed monodisperse ellipse packings for (top row, left to right) $\alpha = 1$, $2$, $3$ $4$, and (bottom row, left to right) $\alpha = 5$, $6$, $8$, $10$.}
\label{fig:snapsmono}
\end{figure*}

\begin{figure*}[htbp]
\includegraphics[width=7in]{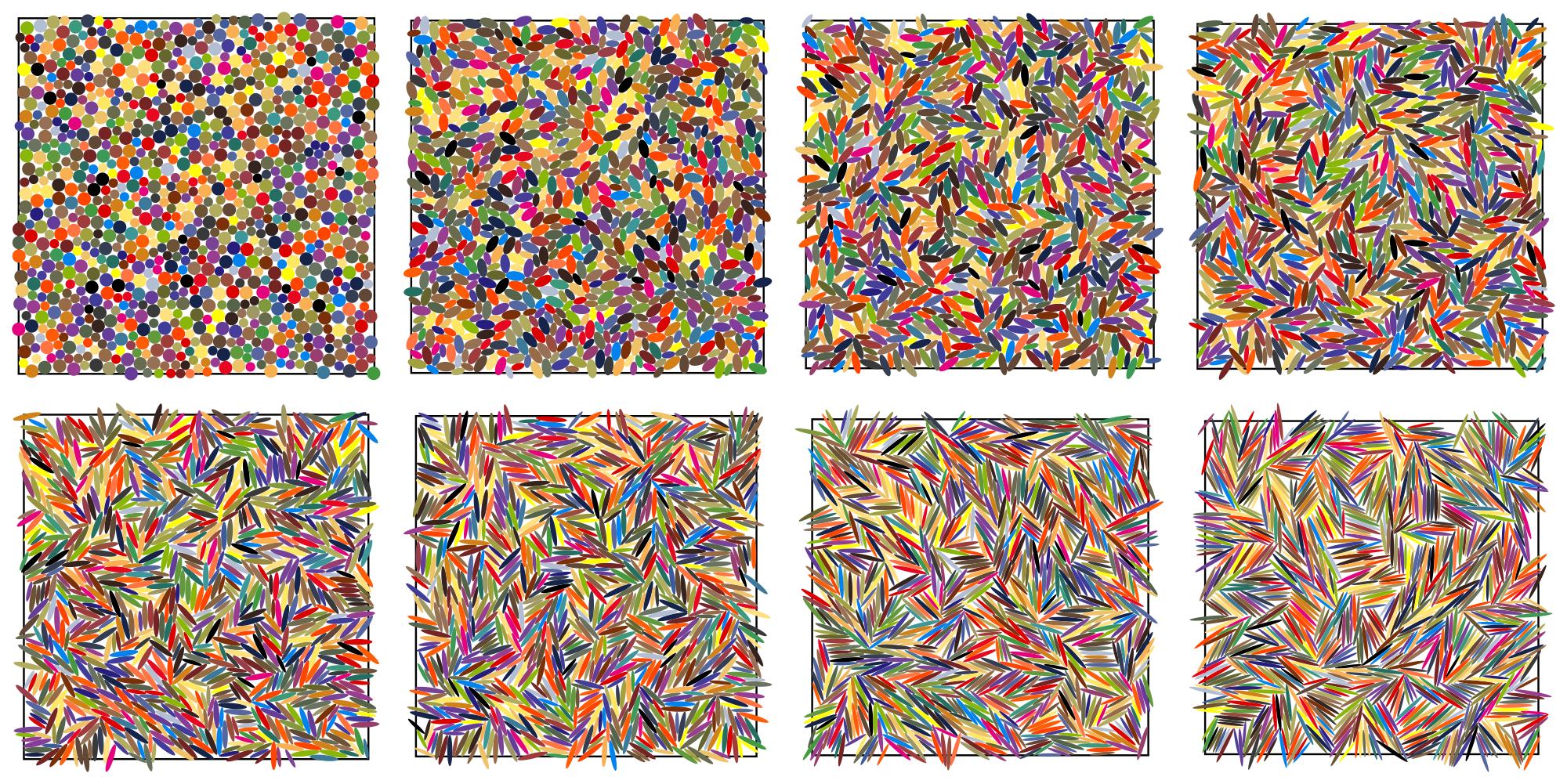}
\caption{Snapshots of jammed 50:50 1:1.4 bidisperse ellipse packings for (top row, left to right) $\alpha = 1$,  $2$, $3$, $4$, and (bottom row, left to right) $\alpha = 5$, $6$, $8$, $10$.}
\label{fig:snapsbi}
\end{figure*}

Figures \ref{fig:snapsmono}-\ref{fig:snapsbi} respectively show snapshots of monodisperse and bidisperse jammed ellipse packings with  $\alpha = 1$, $2$, $3$, $4$, $5$, $6$, $8$, and $10$.
Continuously-polydisperse packings are not shown here because they are very similar to their bidisperse counterparts.
Results for $\alpha = 1$ are entirely as expected from Refs.\ \cite{lubachevsky91,speedy98b,ohern03}: bidisperse packings are disordered and approximately isostatic, while monodisperse disk packings are denser and exhibit long-range triangular-crystalline order interrupted by vacancies and line defects.
For $\alpha = 2$ and $3$, results are consistent with Refs.\  \cite{donev04,donev07,mailman09,delaney05,schreck10,vanderwerf18}.
Visual inspection suggests they  the monodisperse packings are somewhat more ordered than their bidisperse counterparts, but the nature of any such differences is not immediately clear.

Local nematic precursor domains comparable to those observed in experiments on ellipsoidal colloids\cite{zheng11,zheng14,mishra13,roller20,roller21} become increasingly apparent as $\alpha$ increases beyond $\sim 3$.
The domains formed by monodisperse systems appear slightly more ordered than those formed by their bidisperse counterparts, but again the nature of any differences in their ordering is unclear from visual inspection alone.
For $\alpha >\sim 6$, systems form well-defined, mostly-single-layer lamellae.
In contrast to the nearly randomly oriented nematic precursors for $3 <\sim \alpha <\sim 5$, neighboring lamellae are increasingly oriented perpendicularly to each other.
This structure, which is reminiscent of ``checkerboard''-like phases (e.g.\ the high-density disordered equilibrium phase formed by hard rods on a lattice \cite{shah22}), is more prominent for monodisperse systems.
Notably,  the incompatible orientation of neighboring lamellae gives rise to increasingly large voids that cannot be filled because rotations of the surrounding particles (which could otherwise lead to further increases in $\phi$) are blocked by other particles; this mechanism leads to the well-known $1/\alpha$ scaling of $\phi_{\rm J}$ in the large-$\alpha$ limit.\cite{philipse96,desmond06}

\subsection{Measures of local positional-orientational order}

\begin{figure}[htbp]
\includegraphics[width=3in]{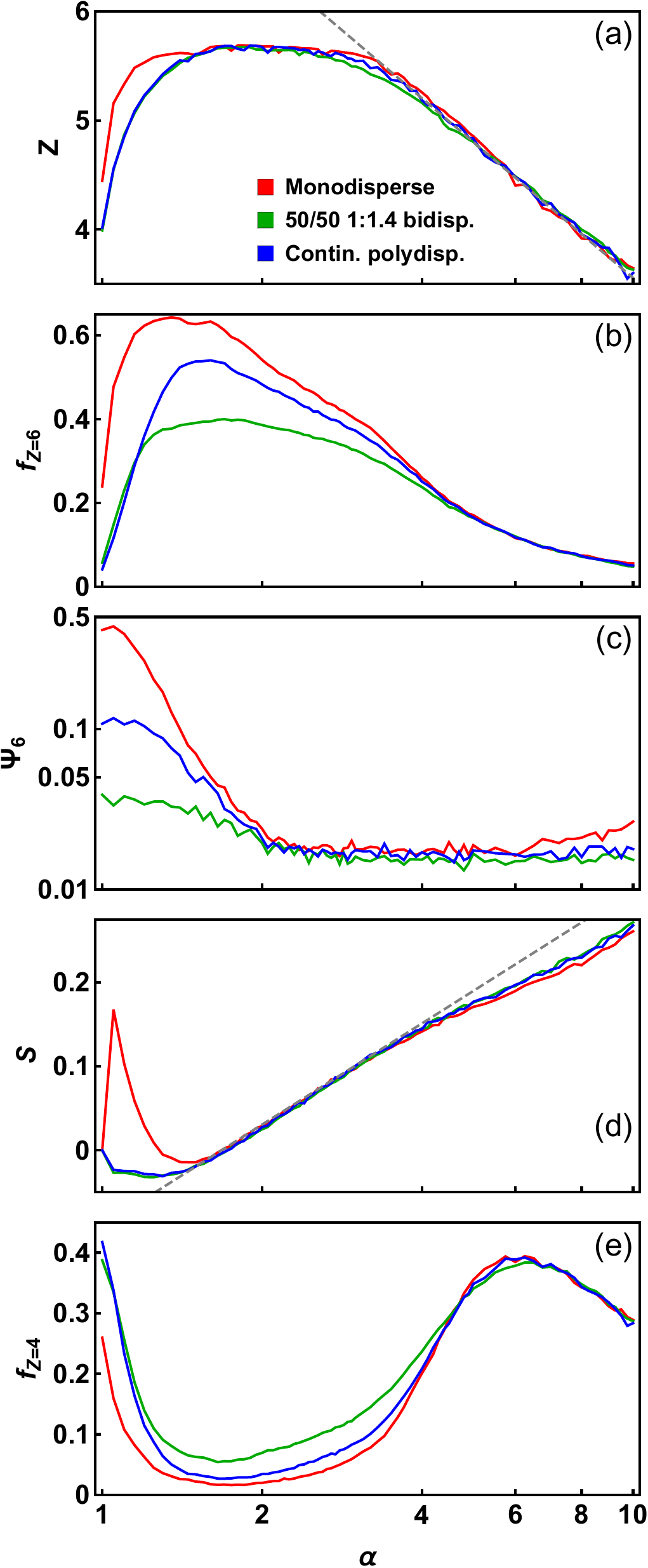}
\caption{Local order parameters for jammed ellipse packings.  All quantities plotted above are defined in  Section \ref{sec:methods}.  Dashed lines in panels (a) and (d) respectively indicate $Z = 7.7-1.8\ln(\alpha)$ and $S = .174\ln(\alpha) - .09$.}
\label{fig:stdops}
\end{figure}

Next, to better understand these variations in local structure, we examine how the structural metrics discussed in Section \ref{sec:methods} vary with $\alpha$.
Figure \ref{fig:stdops}(a) shows results for the  coordination number $Z_{\rm J}$.
Results for small $\alpha$ are consistent with previous work,\cite{donev07,mailman09} showing both the characteristic square-root singularity [$Z_{\rm J}(\alpha) - Z_{\rm J}(1) \propto \sqrt{\alpha - 1}$] for $\alpha - 1 \ll 1$] and convergence towards a plateau at moderate hypostaticity [$Z_{\rm J} = Z_{\rm iso} - \epsilon$ with $\epsilon = 0.3-0.4$] for $1.5 <\sim \alpha <\sim 2.5$.
For $\alpha >\sim 4$, however, $Z_{\rm J}$ drops roughly logarithmically:\ $Z_{\rm J} = Z_0 - b\ln(\alpha)$, with a slightly-dispersity-dependent $Z_0$, and $b \simeq 1.8$. 
This drop in $Z_{\rm J}$ was not observed in previous simulations of ellipse jamming (only one of which\cite{delaney05} reported $Z_{\rm J}$ for $\alpha > 2.5$), but comparable decreases have been reported for rigid-rod-like and semiflexible polymers.\cite{rodney05,hoy17}
Below, we will show that this decrease in $Z_{\rm J}$ is directly associated with an increase in low-coordinated rattler particles trapped inside locally nematic regions.

Figure \ref{fig:stdops}(b) shows the fraction $f_{\rm Z = 6}$ of particles that have exactly six contacts.
For all particle dispersities, the $f_{\rm Z = 6}(\alpha)$ curves have broad peaks centered at $\alpha \simeq \alpha_{\rm max}$.
In other words, maximizing $\phi_{\rm J}$ closely corresponds to maximizing the number of 6-coordinated particles.
Monodisperse particles have both larger $\phi_{\rm J}$ and larger  $f_{\rm Z = 6}$ than their polydisperse counterparts for $\alpha < \alpha_{\rm max}$, owing largely to their greater apparent crystallinity.
Results for different particle dispersities merge for $\alpha >\sim 5$; few 6-coordinated particles are present in these systems.

Since the densest packings have the most six-coordinated particles, a natural followup question is: are they also the most locally hexatically ordered?
Results for $\Psi_6(\alpha)$ [Figure \ref{fig:stdops}(c)] suggests that the answer is: yes, but only when comparing results for different particle dispersities at the same $\alpha$ for $\alpha - 1 \ll 1$.
Intriguingly, $\Psi_6$ is actually slightly larger for $\alpha = 1.05$ than for $\alpha = 1$, suggesting that for increasing  $\alpha - 1 \ll 1$ the ability of particles to rotate away from contacts enhances their ability to hexatically order even as they become more anisotopic. 
Results for larger $\alpha$ show that $\Psi_6$ steadily declines with increasing $\alpha$ for $\alpha >\sim 1.2$ and is minimal for all $\alpha >\sim 2$.
While $\Psi_6$ will decrease with increasing $\alpha$ even for a uniaxially stretched triangular lattice (the densest possible monodisperse ellipse packing, which has $\phi = \phi_{\rm xtal}$ for all $\alpha$\cite{toth50}), the actual decrease shown in Fig. 4(c) is substantially faster than would occur for such a lattice.

Sharper insights into the evolution of jammed ellipse packings' structure are obtained by examining other metrics.
Figure \ref{fig:stdops}(d) shows that the nematic order parameter $S$ is strongly dispersity-dependent for small $\alpha$ but nearly dispersity-independent for $\alpha >\sim 1.8$.
The prominent small-$\alpha$ peak for monodisperse systems coincides with the abovementioned peak in their $\Psi_6$; in the jammed packings for $\alpha <\sim \alpha_{\rm max} = 1.3$, many particles have 6 contacts \textit{and} are aligned with their nearest neighbors.
These regions resemble  a uniaxially stretched triangular lattice.
For bidisperse and continuously-polydisperse systems, $S$ actually becomes negative for $1 < \alpha <\sim 1.8$ because tip-side contacts are favored over side-side contacts in these systems.
For $\alpha >\sim 1.8$, all systems' $S$ increases roughly logarithmically with $\alpha$, with a crossover to a slightly slower rate of increase that corresponds to the emergence of well-defined locally nematic domains over the range $4 <\sim \alpha <\sim 6$.
The beginning of this crossover regime roughly coincides with the end of the $Z_{\rm J} = Z_{\rm iso} - \epsilon$ plateaus shown in Fig.\ \ref{fig:stdops}(a).
In other words, formation of increasingly-well-defined locally-nematic regions within jammed states causes their $Z_{\rm J}$ to drop.

This effect can be further elucidated by examining  $f_{\rm Z = 4}(\alpha)$ [Fig.\ \ref{fig:stdops}(e)].
For $\alpha <\sim 4$,  $f_{\rm Z = 4}$ mirrors  $f_{\rm Z = 6}$.  
Next $f_{\rm Z = 4}$ increases sharply as local nematic domains emerge, reaching a peak at approximately the end of the $S$'s crossover regime, i.e.\ at $\alpha \simeq 6$.
Finally. for $\alpha >\sim 6$, $f_{\rm Z = 4}$ drops again.
These trends can be explained as follows: $f_{\rm Z = 4}$ increases sharply as local nematic domains emerge because (as shown in Figs.\ \ref{fig:snapsmono}-\ref{fig:snapsbi}) these domains lend themselves to $Z = 4$ configurations where ellipses are trapped by one parallel-aligned neighbor on either side and one perpendicularly-aligned neighbor on either end. 
As $\alpha$ continues to increase, the increasing number of rattlers with $Z < 4$, leads to decreasing $f_{\rm Z = 4}$.

\begin{figure}[htbp]
\includegraphics[width=3in]{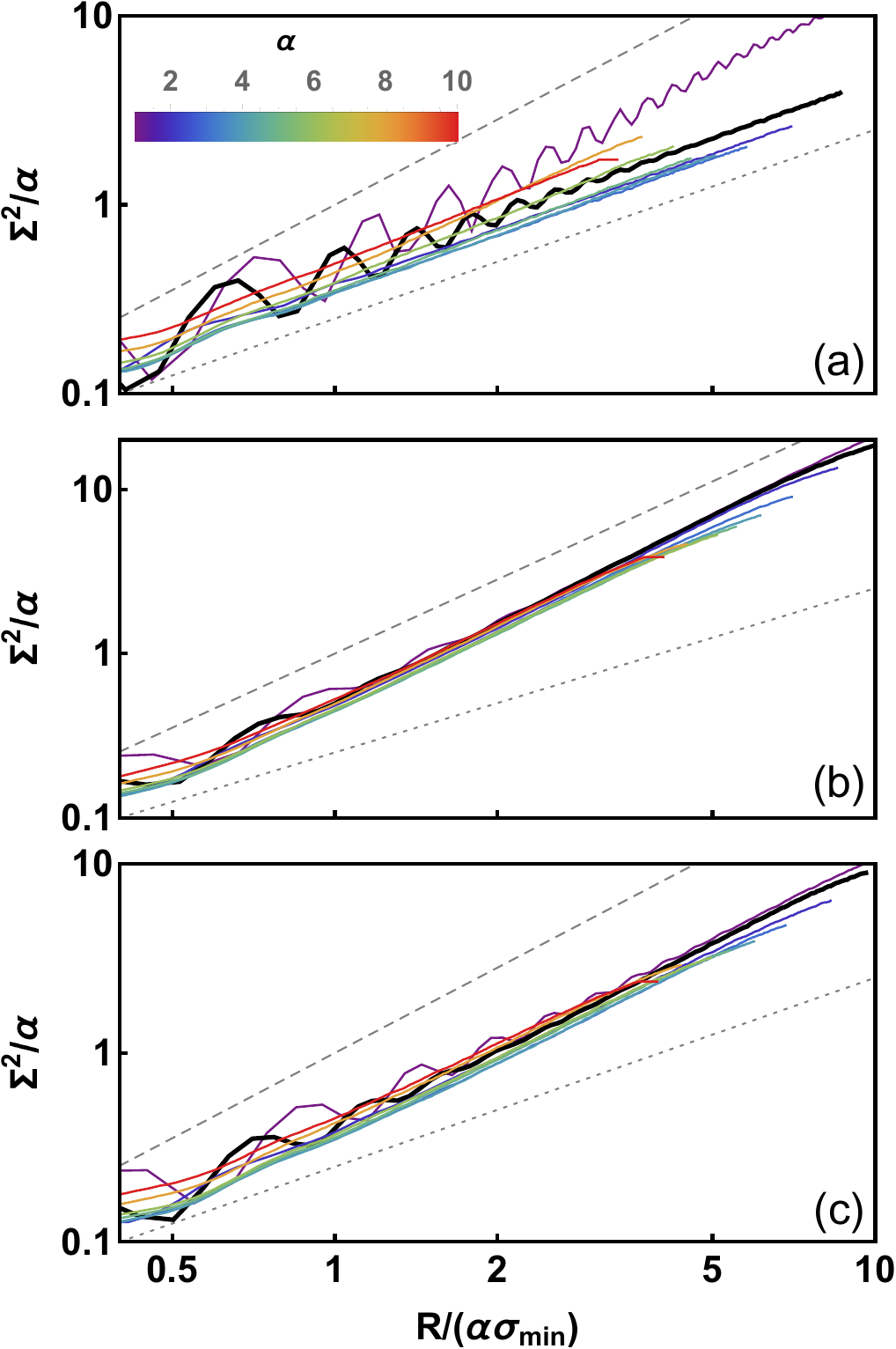}
\caption{Uniformity of jammed ellipse packings.  Panels (a-c) respectively show results for monodisperse, bidisperse, and continuously-polydisperse systems.  Results for the $\alpha$ highlighted in Figs.\ \ref{fig:snapsmono}-\ref{fig:snapsbi} are shown in the colors indicated on the legend, while results for $\alpha = \alpha_{\rm max}$ are shown in black.  The dotted and dashed curves respectively indicate $\Sigma^2/\alpha = 0.17(R/\alpha)$ and $\Sigma^2/\alpha = (R/\alpha)^{3/2}$.  Here $\sigma_{\rm min}$ is the minor-axis length of the smallest particles for the given $\alpha$ and dispersity.}
\label{fig:numvar}
\end{figure}

One might expect that systems with $\alpha \simeq \alpha_{\rm max}$ are maximally dense because they are maximally uniform, and (as will be illustrated below) visual inspection suggests that this is indeed the case.
However, as shown in Figure \ref{fig:numvar}, except for the small-$R$ oscillations associated with the locally crystalline order of monodisperse small-$\alpha$ packings,  $\Sigma^2(R)$ results for all particle dispersities and all $\alpha$ are \textit{qualitatively} very similar, and indeed results for systems of fixed dispersity nearly collapse when  $\Sigma^2/\alpha$ is plotted vs.\ $R/\alpha$.
A completely random arrangement of ellipses would have $y = 1$, i.e.\  $\langle n(R) \rangle \propto \alpha (R/\alpha)^2$ and $\Sigma^2(R) \sim \langle n(R) \rangle  \propto \alpha (R/\alpha)^2$, while a crystalline or quasicrystalline ellipse packing would produce $\Sigma^2(R)  \sim \langle n(R) \rangle^{1/2} \propto \alpha^{1/2} (R/\alpha)$.\cite{torquato03}
While our $N = 1000$ packings are too small to rigorously evaluate the large-$R$ asymptotic scalings of their $\Sigma^2(R)$, we find that they have $\Sigma^2 \sim \alpha (R/\alpha)^y$ with $1 < y <\sim 3/2$ over the range of $R/\alpha$ that allow good statistical sampling.
The imperfect collapses of the data in panels (a-c) indicate that the growth of $\Sigma^2$ with $\alpha$ [at fixed $(R/\alpha)$] is slightly sublinear in $\alpha$ for $\alpha <\sim 6$ and supralinear in $\alpha$ for $\alpha >\sim 6$.
The crossover between these growth regimes reflects the change from (i) a net suppression of density fluctuations for $\alpha <\sim 6$ (compared to those that would be present in completely random packings) by hard-particle excluded-volume constraints, to (ii) a net enhancement of density fluctuations  for $\alpha >\sim 6$ that reflects the increasing contrast between the high-density regions inside the nematic domains and the low-density regions at the boundaries between them.

\begin{figure}[htbp]
\includegraphics[width=3.375in]{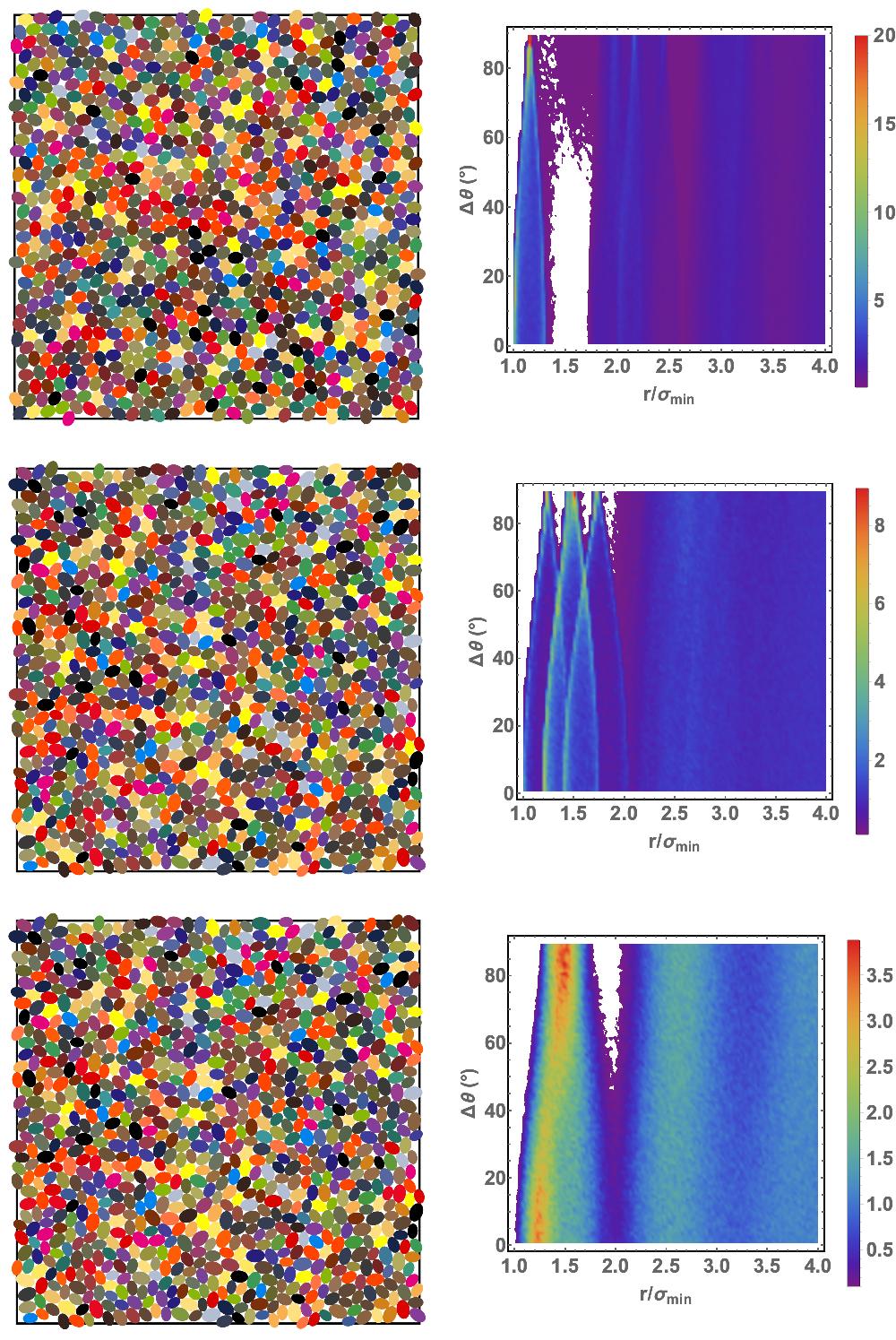}
\caption{Snapshots (left panels) and $g(r,\Delta\theta)$ (right panels) for the densest jammed states for each particle-dispersity category.  Top panels show monodisperse systems with $\alpha = 1.3$, middle panels show 50:50 1:1.4 bidisperse systems with $\alpha = 1.45$, and bottom panels show continuously-polydisperse systems with $\alpha = 1.45$.  Colors are assigned only to regions with $g(r,\Delta\theta) > 0.1$, so both the sterically forbidden and kinetically suppressed regions are shown in white.}
\label{fig:excludedregions}
\end{figure}

While the dataset presented above provides many insights, it fails to conclusively specify what (other than higher $f_{\rm Z = 6}$) distinguishes the densest packings from their lower-$\phi_{\rm J}$ counterparts.  
We now show that this can be done by examining positional-orientational correlations.
Figure \ref{fig:excludedregions} shows representative snapshots and ensemble-averaged $g(r.\Delta\theta)$ for systems with $\alpha = \alpha_{\rm max}$.

The monodisperse packing plainly has a mid-to-long-range crystalline order that superficially resembles that of the triangular lattice.
Nearly all particles have exactly six nearest neighbors that are easily discernible through visual inspection, even though many particles have $Z < 6$ (i.e.\ fewer than six \textit{contacts}).
However, in contrast to the densest crystalline ellipse packing (in which all ellipses are oriented in the same direction and thus have $\Delta\theta = 0$), these nearest-neighbor particles exhibit a wide range of $\Delta\theta$.
Tip-to-side contacts are heavily favored, with $g(r,\Delta\theta) > 30$ in the limit corresponding to perpendicularly-oriented contacting ellipses, i.e.\ $r/\sigma_{\rm min} \to (\alpha+1)/2$ and $\Delta\theta \to 90^\circ$.
At the same time, $g(r,\Delta\theta) < .01$ for certain  ($r,\Delta\theta$) that are sterically allowed (i.e.\ compatible with 2-body hard-particle impenetrability constraints) yet are strongly suppressed by collective many-body effects.
The corresponding minima in $g(r,\Delta\theta)$ are both broad and deep: for example, $g(r,\Delta\theta) < .1$ for all $1.4 < r/\sigma_{\rm min} < 1.7$ with $\Delta \theta \ll 90^\circ$.

The same trends are present for bidisperse and continuously-polydisperse systems even through their $g(r,\Delta\theta)$ are qualitatively different.
More specifically, although  increasing particle dispersity changes the locations of  $g(r,\Delta\theta)$'s extrema, reduces the height and increases the width of its maxima, and reduces both the depth and width of its minima, these minima remain both broad and deep.
We refer to the ranges of ($r,\Delta\theta$) that are sterically allowed yet have  $g(r,\Delta\theta) < 0.1$ as ``kinetically suppressed'' because the various collective many-body ordering processes that occur during dynamic compression make these configurations at least an order of magnitude less likely in the final jammed packings than they would be in  completely disordered packings (i.e.\ ideal gases) with the same $\phi$.
Critically, for all three particle dispersities, the kinetically suppressed regions are largest for $\alpha \simeq \alpha_{\rm max}$, and are absent for systems with $\phi_{\rm J} \leq \phi_{\rm J, disks}$.

Comparing Fig.\  \ref{fig:excludedregions} as well as $g(r,\Delta\theta)$ results for other $\alpha$ (not shown here) to the results presented above shows that large kinetically suppressed regions are present in systems where most particles have six clearly-distinguishable nearest neighbors, whether they actually \textit{contact} all of these neighbors or not.
Nearest-neighbor shells including six members are ``full;'' they prevent any other particles from achieving close proximity, and they do so in a highly $\alpha$- and $\Delta\theta$-dependent way.
As a consequence, systems in which most particles' nearest-neighbor shells are full have richly structured $g(r,\Delta\theta)$ with large kinetically suppressed regions.
These regions are not present in saturated RSA ellipse packings,\cite{abritta22} which suggests that they arise during the later stages of compression, i.e.\ over the range $\phi_{\rm s}(\alpha) <\sim \phi < \phi_{\rm J}(\alpha)$.

\subsection{Comparison to RSA packings}

For a wide variety of particle shapes, complex liquid-state dynamics are expected for packing fractions in the range $\phi_{\rm o}(\alpha) < \phi < \phi_{\rm g}^{\rm trans}(\alpha)$, where $\phi_{\rm o}(\alpha)$ is the ``onset'' density.\cite{sastry98,chaudhuri10} 
In hard-ellipse liquids, onset and translational-rotational decoupling\cite{chong09} have been associated with the emergence of  unstable nematic-like regions with a mean lifetime $\tau_{\rm nem}$ that exceeds the characteristic relaxation time $\tau_{\rm 0}$ for translational diffusion.\cite{davatolhagh12}
Measurement of the ratios $\phi_{\rm g}^{\rm trans}(\alpha)/\phi_{\rm g}^{\rm rot}(\alpha)$, $\phi_{\rm g}^{\rm trans}(\alpha)/\phi_{\rm o}(\alpha)$ and $\phi_{\rm g}^{\rm rot}(\alpha)/\phi_{\rm o}(\alpha)$ for various shapes over a wide range of $\alpha$ could provide additional valuable insights into these dynamics, but evaluating these quantities is computationally expensive.\cite{pfleiderer08,shen12}  
An alternative approach that should provide at least some of the same insights is to measure the ratio $\phi_{\rm J}(\alpha)/\phi_{\rm s}(\alpha)$, where the RSA density $\phi_{\rm s}(\alpha)$ is the maximum density at which impenetrable particles of aspect ratio $\alpha$ can be packed under a protocol that sequentially inserts them with random positions and orientations.
This ratio of fundamental interest because it indicates how much packing efficiency particles can gain via cooperative translations and rotations during the later stages of compression, i.e.\ over the range $\phi_s(\alpha) < \phi < \phi_J(\alpha)$.
Surprisingly, to the best of our knowledge, no previous studies have systematically examined $\phi_{\rm J}(\alpha)/\phi_{\rm s}(\alpha)$ for ellipses, ellipsoids, or other comparable 2D or 3D convex shapes.

\begin{figure}[htbp]
\includegraphics[width=3in]{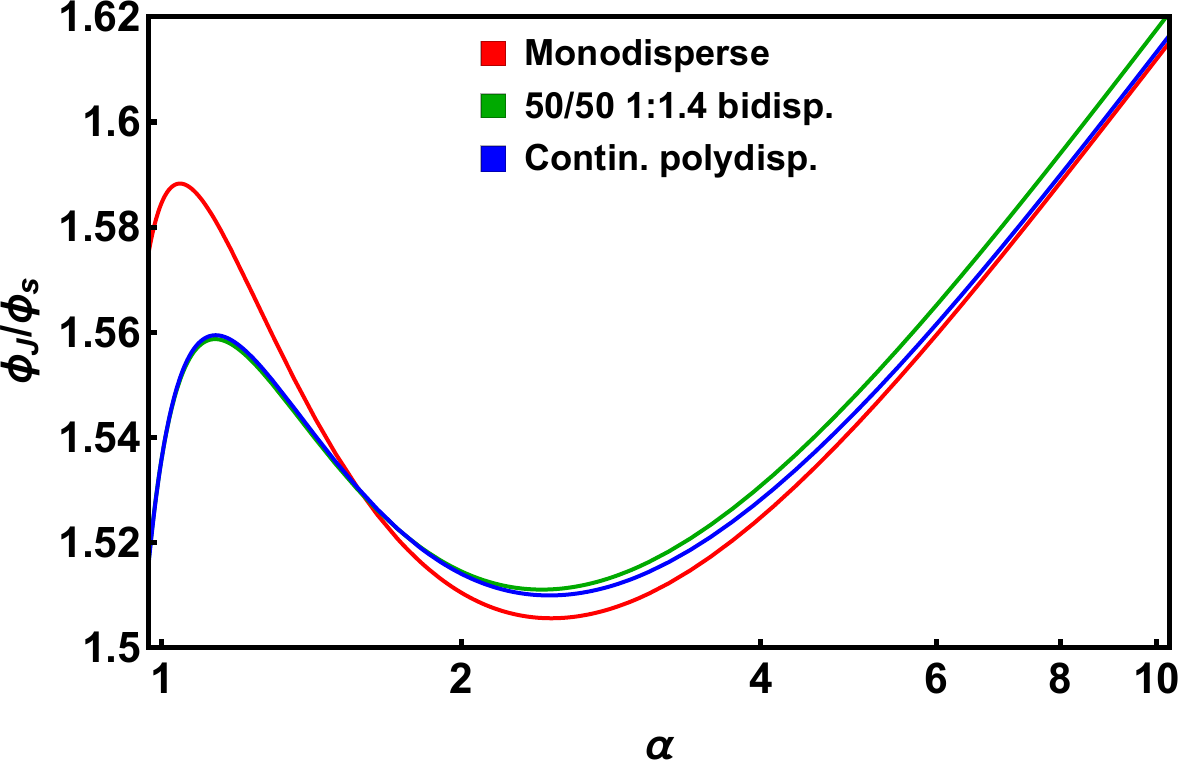}
\caption{Ratio of the jamming densities $\phi_{\rm J}^{\rm x}(\alpha)$ to the saturated RSA packing densities $\phi_{\rm s}(\alpha)$ \cite{abritta22} of monodisperse ellipses.}
\label{fig:phiJphiSratio}
\end{figure}

Remarkably, our expressions for  $\phi_{\rm J}^{\rm x}(\alpha)$ (Eqs.\  \ref{eq:phiJmono}-\ref{eq:phiJcontin}) have the same functional form as one that predicts monodisperse ellipses' $\phi_{\rm s}(\alpha)$ to within $\sim 0.1\%$ over the same range of $\alpha$ ($1 \leq \alpha \leq 10$) considered here:\cite{abritta22}
\begin{equation}
\phi_{\rm s}(\alpha) = \phi_{\rm s, disks} \times \displaystyle\frac{1 + \frac{3}{8}\ln(\alpha) +  \frac{17}{25}(\alpha-1)}{1 + \frac{80}{99}(\alpha - 1) + \frac{1}{96}(\alpha-1)^2},
\label{eq:phis}
\end{equation}
where $ \phi_{\rm s, disks} = .54707$.\cite{zhang13}  
As shown in Figure \ref{fig:phiJphiSratio}, in our bidisperse and continuously-polydisperse systems, the ratio $\phi_{\rm J}(\alpha)/\phi_{\rm s}(\alpha)$ stays within $\sim 1\%$ of 1.53 for all $1 \leq \alpha \leq 5$.
$\phi_{\rm J}(\alpha)/\phi_{\rm s}(\alpha)$ is larger for our small-$\alpha$ monodisperse systems, and for all dispersities for $\alpha >\sim 5$. 
In other words, our data indicates that this ratio is almost $\alpha$-independent as long as neither substantial local hexatic order nor substantial local nematic order develops during compression.

\section{Discussion and Conclusions}
\label{sec:conclude}

In this paper, we performed a detailed characterization of jammed ellipse packings over a much wider range of aspect ratios ($1 \leq \alpha \leq 10$) than had previously been attempted.
Our first major goal was to determine $\phi_{\rm J}(\alpha)$ to high precision, for three different particle dispersities: mono-, bi-, and continuously-polydisperse.
After doing so, we found simple analytic formulae (Eqs.\ \ref{eq:phiJmono}-\ref{eq:phiJcontin}) that predict these $\phi_{\rm J}$ to within $<\sim 0.1\%$.
Surprisingly, ellipses' jamming and saturated-RSA packing densities are both quantitatively predicted over entire range of $\alpha$ by a common functional form
\begin{equation}
\displaystyle\frac{\phi_{\rm X}(\alpha)}{\phi_{\rm X}(1)} = \displaystyle\frac{1 + a\ln(\alpha) +  b(\alpha-1)}{1 + c(\alpha - 1) + d(\alpha-1)^2},
\label{eq:analyticphiX}
\end{equation}
where $\phi_{\rm X}$ is the jamming or RSA density (i.e.\ $\phi_{\rm J}$ or $\phi_{\rm s}$) and the coefficients $\{ a, b, c, d \}$ depend on particle dispersity and the packing preparation protocol.
Moreover, the ratio $\phi_{\rm J}(\alpha)/\phi_{\rm s}(\alpha)$ remains almost $\alpha$-independent, suggesting that the amount of extra packing efficiency ellipses can gain via cooperative translations and rotations during the later stages of compression depends only depends only weakly on their anisotropy, as long as neither substantial local hexatic nor substantial local nematic order develops during compression.
Comparison to previous results for other particle types including spherocylinders and strongly-overlapping $n$-mers\cite{ciesla16,marschall18} suggests that Eq.\ \ref{eq:analyticphiX} may be applicable to all convex 2D shapes, with $\{ a, b, c, d \}$ that depend on particles' shape in addition to the factors mentioned above.

Our second major goal was to characterize the local structure of higher-$\alpha$ packings including the local nematic domains found in liquid-glass colloidal suspensions.\cite{zheng11,zheng14,mishra13,roller20, roller21}
Previous studies of ellipse jamming found that $Z_{\rm J}(\alpha)$ plateaus at moderate hypostaticity [$Z_{\rm J} = 6 - \epsilon$ with $\epsilon = 0.3-0.4$ for $1.5 <\sim \alpha <\sim 2.5$],\cite{donev07,mailman09,vanderwerf18} and implied that this plateau extends to $\alpha = \infty$.
However, since these studies did not examine $\alpha$ that were sufficiently large to possess a high-$\phi$ equilibrium nematic phase (e.g.\ $\alpha > 2.4$ for monodisperse ellipses\cite{bautista14b}) and hence only examined nearly-isotropic packings, the question of whether it actually does so had remained open.
Here we found that $Z_{\rm J}$ drops roughly logarithmically [$Z_{\rm J} \simeq Z_0 - b\ln(\alpha)$, with weakly-dispersity-dependent $Z_0$ and $b$] for $\alpha >\sim 3$.
This drop in $Z_{\rm J}$ results largely from an increasing fraction of particles that are trapped inside locally nematic domains by a parallel-oriented neighbor on either side and a perpendicularly-oriented neighbor on either end, and hence have no more than four contacts.
The emergence of comparable particle caging during dynamic compression may help explain the onset of liquid-glass physics in athermal systems.\cite{davatolhagh12}

The final major question we wished to answer in this study was: what structural features distinguish the densest jammed packings from their lower-$\phi_{\rm J}$ counterparts?
Examination of commonly employed structural metrics such as the local nematic order parameter $S$, the Steinhardt-like order parameter $\Psi_6$\cite{bernard11} and the uniformity metric $\Sigma^2(R)$\cite{torquato18b} failed to conclusively answer this question.
Instead we showed that the fraction of particles that have exactly six contacts ($f_{\rm Z = 6}$) is maximized at $\alpha \simeq \alpha_{\rm max}$ for all particle dispersities even though $f_{\rm Z = 6}(\alpha)$ is itself highly dispersity-dependent, and that locally-hyperstatic particles within $\alpha \simeq \alpha_{\rm max}$ packings are far more likely to have six clearly-distinguishable nearest neighbors than their counterparts in systems with $\phi_{\rm J} < \phi_{\rm J, disks}$, even in the absence of substantial local hexatic order.
While it has long been known that nearest-neighbor shells including six members are full and hence prevent any other particles from achieving close proximity to the reference particle, here we showed that they do so in a highly $\alpha$- and $\Delta\theta$-dependent way that (in systems with $\alpha \simeq \alpha_{\rm max}$) leads to richly structured $g(r,\Delta\theta)$ with large kinetically suppressed regions.
In other words, we showed that particles with $\alpha \simeq \alpha_{\rm max}$ develop unusually-well-defined nearest-neighbor shells during compression, for three very different particle dispersities, even through the structure of the shells themselves is highly dispersity-dependent.
We conclude that it is these well-defined shells that allow  $\alpha \simeq \alpha_{\rm max}$ ellipses' $\phi_{\rm J}$ to be substantially higher than disks' $\phi_{\rm J}$ even though their jammed states do not possess longer-range crystalline order.
This conclusion places Donev \textit{et al.}'s argument that ellipses' ability to rotate away from contact allows them to pack more densely than disks\cite{donev04} on a firmer quantitative foundation.

\section*{Conflicts of interest}

There are no conflicts to declare.

\section*{Acknowledgements}

This material is based upon work supported by the National Science Foundation under Grant DMR-2026271.

%\bibliographystyle{rsc}
%\bibliography{glassjamming}

\providecommand*{\mcitethebibliography}{\thebibliography}
\csname @ifundefined\endcsname{endmcitethebibliography}
{\let\endmcitethebibliography\endthebibliography}{}

\end{document}